\documentstyle[graphicx,float]{article}

\begin{document}
\title{Symmetrisation effects on the correlation time delay}
\date{}
\author{Pedro Sancho \\ Centro de L\'aseres Pulsados CLPU \\ Parque Cient\'{\i}fico, 37085 Villamayor, Salamanca, Spain}
\maketitle
\begin{abstract}
We analyze the electronic correlation contribution to the time delay in the photo-ionization of the excited ortho- and para-Helium states. A simple estimation, based on the ionization probability amplitudes, shows that the different form of anti-symmetrising both states can in principle lead to very different values of the correlation time delay. This result illuminates the interplay between electronic correlations and symmetrisation effects in the  attosecond regime, a relation that has been studied in other contexts. Moreover, it suggests the potential of excited states to explore the role of exchange effects in that realm.  
\end{abstract}

\section{Introduction}

The advent of attosecond physics made possible the experimental study of ionization times of atoms and molecules \cite{nat,sci}. Attosecond time-resolved spectroscopy demonstrated that the photo-ionization of atoms and molecules is not an instantaneous process. There is a time delay between the interaction with the light pulse and the electron emission.  More recently, even the dependence of the delay on the electronic correlations present in the Helium atom has been numerically and experimentally quantified  \cite{co1,co2}. 

In the field of light-matter interactions there are many examples of modifications of physical processes by particle identity. For instance, light absorption by atoms is strongly dependent on the symmetrisation of the atomic state \cite{yo1}. We want consider in this paper if a similar dependence can be present in the attosecond regime, where the relevant time scales are much shorter. A well suited scenario to test that idea is the study of time delays in ionization.    

We restrict our considerations to the simplest case where that influence can manifest, the ortho- and para-Helium states. The determination of the ionization time normally considers the ground state of the atom or molecule. For excited states new possibilities emerge. For instance, the spatial and spin parts of the ortho and para states are oppositely symmetrised. We show that, in principle and depending on the actual values of the parameters of the problem, these differences can lead to rather different contributions of the electronic correlations to the time delay in ionization. 

The actual determination of the problem parameters can only be done via experimental measurement or detailed numerical simulation. We shall not try this exact approach. Instead, we only present a simple evaluation of the ionization probabilities, which allows for an estimation of the correlation time delay dependence on the antisymmetrisation form of each state. This approximate estimation shows that for many values of the parameters there are large differences between the ortho- and para-Helium correlation time delays. Our aim is only to provide a proof of principle of the potential of excited states to study symmetrisation effects in ionization processes in attosecond physics.  

The analysis presented in the paper reflects the complex relation between the electronic correlations and the symmetrisation process. This relation shows many similarities with the subtle connection between entanglement and particle identity, which we discuss at the end of the paper. 

\section{Ionization of excited states}

In this section we evaluate the ionization probability amplitude of excited states. These amplitudes are necessary to estimate the electronic correlation contribution to the time delay. As it is well-known the state of the two electrons of an excited Helium atom can be antisymmetrised in two different ways, by symmetrising the spatial part and antisymmetrising the spin one, or vice versa (the two parts are in a product form in the full state). We refer to these options as the para- and ortho-Helium states,  corresponding to spin singlet and triplet states. For the first excited states with one of the electrons in the level $1s$ and the other in the $2s$ one, they are:
\begin{equation}
|\Psi_{par}>=N_{par}(|\Psi _{1s2s}>_{12} + |\Psi _{1s2s}>_{21}) |singlet>     
\end{equation}
and
\begin{equation}
|\Psi_{ort}>=N_{ort}(|\Psi _{1s2s}>_{12} - |\Psi _{1s2s}>_{21}) ||triplet>     
\end{equation}
with $|singlet>= (|+>_1|->_2-|->_1|+>_2)/\sqrt 2$ and three different options for the $|triplet>= (|+>_1|->_2+|->_1|+>_2)/\sqrt 2, |+>_1|+>_2$ and $|->_1|->_2$, where $+$ and $-$ denote the two spin components of each electron. The wave functions corresponding to the kets $|\Psi_{1s2s}>_{12}$ and  $|\Psi_{1s2s}>_{21}$ are respectively $\Psi _{1s2s}({\bf x},{\bf y}) $ and $ \Psi _{1s2s}({\bf y},{\bf x})$, with $\bf x$ and $\bf y$ the spatial coordinates of the two electrons ($1$ and $2$). $\Psi_{1s2s}({\bf x},{\bf y})$ is a solution of the Schr\"{o}dinger equation incorporating the two-electron Coulomb interaction and, consequently, representing a non-separable state. Finally, the normalization coefficients are
\begin{equation}
N_{par}=(2+2Re(_{12}<\Psi_{1s2s}|\Psi_{1s2s}>_{21}))^{-1/2}
\end{equation}
and
\begin{equation}
N_{ort}=(2-2Re(_{12}<\Psi_{1s2s}|\Psi_{1s2s}>_{21}))^{-1/2}
\end{equation} 
provided that $|\Psi _{1s2s}>_{12}$ and $|\Psi _{1s2s}>_{21}$ are normalized.

If the Coulomb interaction between the electrons is neglected, the spatial states will be in the non-entangled separable form typical from the central field approximation \cite{bys}:
\begin{equation}
|\bar{\Psi}_{par}>=\frac{1}{\sqrt 2}(|\psi _{1s}>_1|\psi _{2s}>_2 + |\psi _{2s}>_1|\psi _{1s}>_2) |singlet>     
\end{equation}
and
\begin{equation}
|\bar{\Psi}_{ort}>=\frac{1}{\sqrt 2}(|\psi _{1s}>_1|\psi _{2s}>_2 - |\psi _{2s}>_1|\psi _{1s}>_2) |triplet>     
\end{equation}
with $\psi _{1s}$ and $\psi _{2s}$ the one particle states of the $1s$ and $2s$ levels.

Just after preparing the atom in the excited state, we repeat the  standard process (see, for instance, \cite{co1}) in order to measure the ionization time. Note that from the experimental point of view this would demand a very fine temporal tuning between the different light beams (exciting, attosecond shake-up ionizing and laser electric probe field) involved in the process. We recall 
that the electronic correlations only modify the ionization times in a relevant form when the energy of the ionizing photons is high enough to extract one electron and to carry the other to an excited state of the remaining ion $He^+$. When this condition is fulfilled we speak of shake-up ionization and shake-up excited states. We restrict our considerations to this ionization regime.

After the interaction the photo-emitted electron and the one remaining bound to the ion $He^+$, are in the antisymmetrised states 
\begin{equation}
|\Phi _{par}^{{\bf p},nl+}>={\cal N}_{par}(|\Psi^{{\bf p},nl+}>_{12} + |\Psi^{{\bf p},nl+}>_{21}) |singlet>
\end{equation}
and
\begin{equation}
|\Phi _{ort}^{{\bf p},nl+}>={\cal N}_{ort}(|\Psi^{{\bf p},nl+}>_{12} - |\Psi^{{\bf p},nl+}>_{21}) |triplet>
\end{equation}
 $\Psi^{{\bf p},nl+}$ is a non-separable state (the electrons are still interacting at that stage via the Coulomb potential). The label ${\bf p}$ refers to the momentum of the state of the continuum and $nl+$ to the ionic excited state levels. The normalization coefficients are
 \begin{equation}
{\cal N}_i=(2 \pm 2Re(_{12}<\Psi^{{\bf p},nl+}|\Psi^{{\bf p},nl+}>_{21}))^{-1/2}
\end{equation}
with $i=par,ort$ and in the double sign expression the $+$ holds for the para state and the $-$ for the orto one. We assume $|\Psi^{{\bf p},nl+}>_{12}$ and $|\Psi^{{\bf p},nl+}>_{21}$ to be already normalized. Note that the state must retain the para (orto) form after the interaction. This is so because of the non-relativistic treatment used here the interaction with the light does not change the spins, and the singlet and triplet states do not change after the light absorption.
 
The probability amplitude for the ionization can be evaluated using the evolution operator $\hat{U}$ of the complete system (atom plus electromagnetic field of the ionizing light pulse)
\begin{equation}
{\cal M}_i^{{\bf p},nl+}=<\Phi_{i}^{{\bf p},nl+}|_{EM}<0| \hat{U}|1>_{EM}|\Psi _i>
\end{equation}
The electromagnetic states of the ionizing beam previous and subsequent to the interaction are $|1>_{EM}$ and $|0>_{EM}$ (we only consider single absorptions). The above equation can be expressed as 
\begin{eqnarray}
{\cal M}_i^{{\bf p},nl+} = {\cal N}_iN_i (_{12}<\Psi ^{{\bf p},nl+}|_{EM}<0| \hat{U}|1>_{EM}|\Psi_{1s2s}>_{12} + \nonumber \\
_{21}<\Psi ^{{\bf p},nl+}|_{EM}<0| \hat{U}|1>_{EM}|\Psi_{1s2s}>_{21} \pm \nonumber \\
\{_{12}<\Psi ^{{\bf p},nl+}|_{EM}<0| \hat{U}|1>_{EM}|\Psi_{1s2s}>_{21} +  \nonumber \\
_{21}<\Psi ^{{\bf p},nl+}|_{EM}<0| \hat{U}|1>_{EM}|\Psi_{1s2s}>_{12} \} )
\label{eq:die}
\end{eqnarray}
The matrix element is composed of four terms. They represent the four alternatives for the transition:

(i) Electron labelled $1$ initially in state $1s$ being emitted to the continuum, and electron $2$ evolving from $2s$ to $nl+$.

(ii) Electron $1$: $2s \rightarrow nl+$. Electron $2$: $1s \rightarrow $ continuum.  

(iii) Electron $1$: $2s \rightarrow $ continuum. Electron $2$: $1s \rightarrow nl+$.  

(iv) Electron $1$: $1s \rightarrow nl+$. Electron $2$: $2s \rightarrow $ continuum.  

Moreover, there is an additional $\pi$ phase between the two last and the two first terms in the orto case. When evaluating the ionization probabilities, half of the interference terms between different terms will have different signs in the orto and para cases. These differences can lead, in principle, to different ionization behaviours.

\section{Correlation time delay}

The contribution to the time delay of the electronic correlation is given by
\begin{equation}
\tau _{cor}= \frac{1}{\omega _L} \arctan \left(  d_{eff} \frac{\omega _L}{p_0} \right) 
\end{equation}
with $\omega _L$ the laser frequency, $p_0$ the initial momentum of the photo-electron and $d_{eff}$ the effective dipole momentum of the ion in the shake-up state. This effective dipole interacts with the outgoing electron, inducing the additional time shift of the leaving electron \cite{co1}. 

The magnitude of the effective dipole can be evaluated taking into account that the dominant contribution is that of the shake-up $nl+=2s+$ and $nl+=2p+$ states of the ion. If we express this shake-up state as the superposition $|\phi >_{n+=2}=c_{2s+}|\phi _{2s+}>+c_{2p+}|\phi _{2p+}>$, with  $c_{2s+}$ and $c_{2p+}$ the amplitudes or populations of the two sub-states, the effective dipole will be given by the equation
\begin{equation}
d_{eff}^{n+=2}=d_0^{n+=2} \cos (\Delta \varphi) \frac{|c_{2s+}/c_{2p+}|}{1+|c_{2s+}/c_{2p+}|^2}
\end{equation}
where $d_0^{n+=2}$ is the maximum value of the dipole moment in that shell, and $\Delta \varphi = \varphi _{2s+} - \varphi _{2p+}$ is the relative phase between the two amplitudes.

In our approach these amplitudes correspond to the probability amplitudes for the transition from the initial state to the shake-up excited state of the ion, and simultaneously the other electron to the continuum, and consequently can be approximated
  by the ${\cal M}_{par}^{nl+}$ and ${\cal M}_{ort}^{nl+}$ amplitudes.

Being the population amplitudes and the relative phases different in the para and ortho cases we must expect the effective dipole moments and, consequently the $\tau_{cor}$, also to be different. A more quantitative characterization of these differences would demand an extensive numerical calculation. Instead, we shall present in the next section a graphical representation of the  $\tau_{cor}$ for some estimated values. 

\section{Graphical representation}

 In order to evaluate $\tau _{cor}$ we must determine $d_{eff}$ and $p_0$ ($\omega _L$ is a free parameter of the arrangement). The problem can be simplified by expressing the equations in terms of the experimental values $\tau _{cor}^{exp}$ and $d_{eff}^{exp}$, that is, of the values obtained in \cite{co1}, and making the natural simplifying assumption that the initial momenta of the photo-electrons are very similar to those observed in \cite{co1}:
\begin{equation}
\tau _{cor} \approx \frac{1}{\omega _L}\arctan \left( \frac{d_{eff}}{d_{eff}^{exp}} \tan (\omega _L \tau _{cor}^{exp}) \right)  
\end{equation} 
The value of the laser frequency is $2\pi /\omega_L=2.7 fs$ and that of the experimental dipole $d_{eff}^{exp}=0.32$ atomic units. These values correspond to attosecond beams with 97 eV photon energy, for which $|c_{2s+}/c_{2p+}|=1.1$ and $\Delta \varphi =1.35$ rad.  

Introducing the notation ${\cal M}_i^j(k)=R_i^j(k) \exp (\varphi _i^j(k))$, with $i=d , c$, $j=2s+,2p+$ and $k=1,..,4$ (the four terms in Eq. (\ref{eq:die})), we can express ${\cal M}_{ort}^j$ and ${\cal M}_{par}^j$ in terms of the  $R_i^j$'s and $\varphi _i^j$'s. We use in Fig. (1) the values $|R_i^{2s+}(k)|=1.05, 0.9,1,1.1$, $|R_i^{2p+}(k)|=1,0.9,1.1,1.15$, $\varphi _i^{2s+}(1)=0$, $\varphi _i^{2s+}(2)=1$, $\varphi _i^{2s+}(3)=2$ and $\varphi _i^{2p+}(k)=3.1,0.2,3,3.1$. On the other hand, $\varphi _i^{2s+}(4)$ is the independent variable. This choice is arbitrary, but with values of the modulus close to those experimentally observed in \cite{co1}.

In Fig. (1) we represent $\tau _{cor}$ as a function of $\varphi _4^{2s+}$. We observe for different $\varphi _4^{2s+}$ large variations of the values of $\tau _{cor}$ in each curve. They oscillate in a very long interval. For most values of the phase they are also long when compared to the correlation time delay of the ground state, that is, to $\tau_{cor}^{exp}$. 
\begin{figure}[H]
\center
\includegraphics[width=13cm,height=7cm]{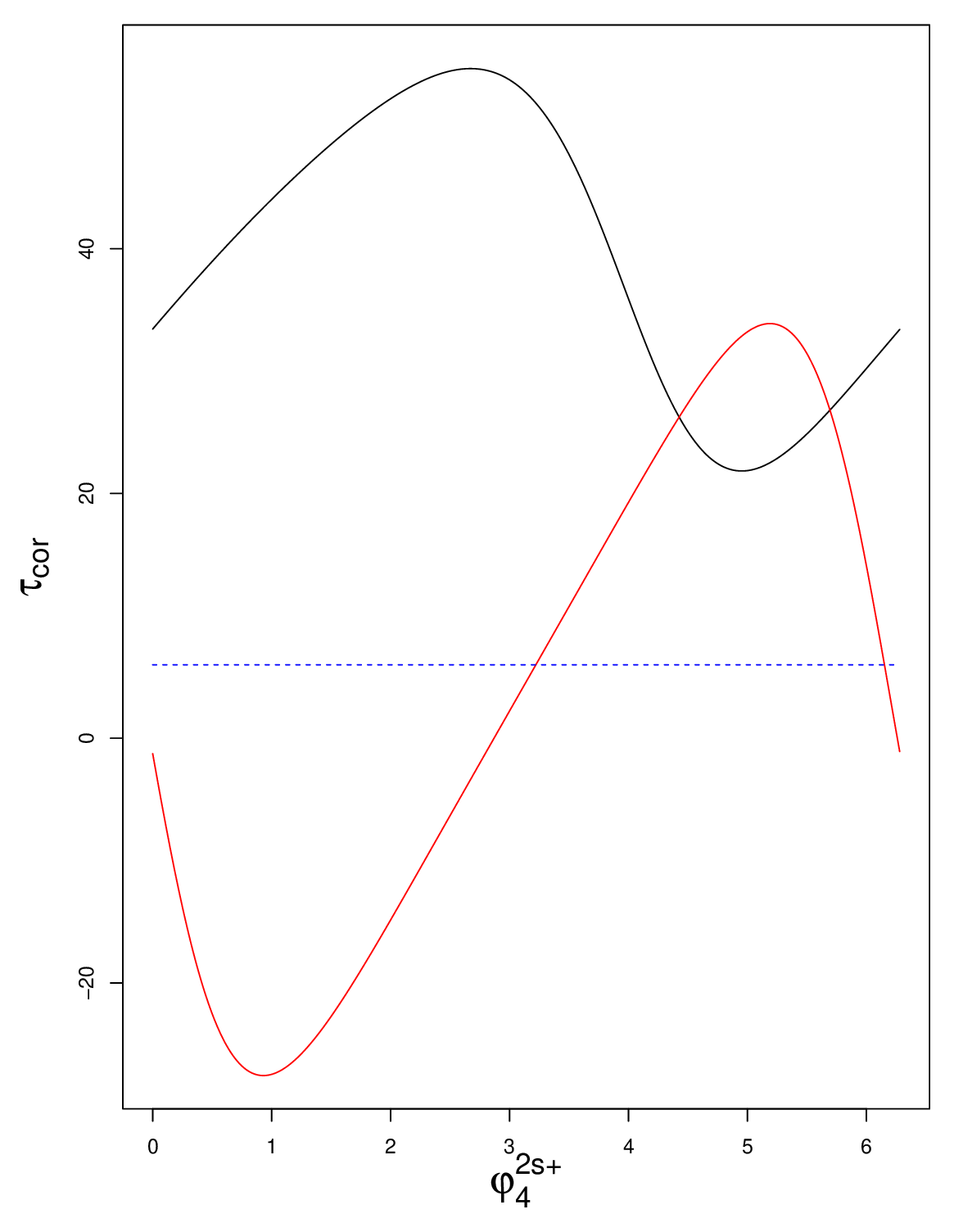}
\caption{Representation of $\tau _{cor}$ in as ($10^{-18}$s) versus $\varphi _4^{2s+}$ in radians. The black and red curves correspond respectively to the para and ortho initial states and the blue dashed one represents the value of $\tau_{cor}^{exp}$.  }
\end{figure}
The differences between the orto and para curves is large. In this example the $\tau _{cor}$ can even have opposite signs. The correlation time delays can have positive and negative values, that is, they can increase or decrease the temporal delay effects associated with the other processes involved (the Eisenbud-Wigner-Smith time shift and the Coulomb-laser coupling \cite{co1}).

We have studied many other configurations with different values of the parameters. A general characteristic is that the curves are strongly dependent on the values of the parameters, although maintaining in general two properties, large differences between the ortho and para initial states, and between excited and ground ones.

\section{Discussion}

We have studied the contribution of the electronic correlations to the time delay in photo-ionization of excited states. The advantage of using these states, instead of the ground one, is that different forms of antisymmetrisation can be considered in the same system. Our results suggest that for some values of the problem parameters there are large differences in the above contributions.

The exchange effects can be viewed as the interference of the alternative paths associated with the different labels of the identical particles. For the ground state of the Helium atom there is only one form to antisymmetrise it. However, for the excited states we have the ortho and para options. These options lead to different forms of superposing the terms associated with the two alternative paths and, consequently, to different interference effects. The interference framework provides an intuitive view of the differences between para- and ortho-Helium.    

We remark that our approach does not provide conclusive results. It does not give exact values for the delays, but a first approximation. Only extensive numerical simulations or experimental measurements would provide these reliable values. Nevertheless, the partial results obtained here show the potential of excited states in the problem.

The analysis presented here also shows the complex interplay between electronic correlations and symmetrisation effects in attosecond dynamics. This relation is reminiscent in many aspects of the entanglement-identity interplay in other regimes. Although entanglement is a controversial subject for identical particles \cite{tri}, the connection with the exchange effects has been considered in the literature. For instance, in \cite{yo2} the complex dynamics of light absorption by identical atoms in entangled states was discussed. Large differences between boson and fermion atoms, and for both large departures with respect to the behaviour of distinguishable systems were found. It would be interesting to compare both types of effects in order to see if they are similar or dependent on the involved time scale.


\begin{thebibliography}{99}
\bibitem{nat} M. Hentschel, R. Kienberger, {\it et al}, Nature {\bf 414} 509 (2001)
\bibitem{sci} M. Schultze, M. Fiess, {\it et al}, Science {\bf 328} 1658 (2010)
\bibitem{co1} M. Ossiander, F. Siegrist, {\it et al}, Nat. Phys. {\bf 13} 280 (2017) 
\bibitem{co2} R. Pazourek, S. Nagele, J. Feist, J. Burgd\"{o}rfer,  Phys. Rev. Lett. {\bf 108} 163001 (2012)
\bibitem{yo1} P. Sancho, Ann. Phys. {\bf 336} 482 (2013) 
\bibitem{bys} S. Wieder, {\it The Foundations of Quantum Theory} (Academic Press, 1973)
\bibitem{tri} F. Benatti, R. Floreanini, F. Franchini, U. Marzolino, Phys. Rep. {\bf 878} 1 (2020)
\bibitem{yo2} P. Sancho, Ann. Phys. {\bf 421} 168264 (2020)
\end{thebibliography}
\end{document}